\documentclass{aastex631}

\usepackage{amsmath}
\usepackage{multirow}

\begin{document}

\title{Long-term evolution of Sco X-1: implications for the current spin frequency and ellipticity of the neutron star}

\correspondingauthor{Abhijnan Kar}

\author{Abhijnan Kar}
\affiliation{Department of Physical Sciences, Indian Institute of Science Education and Research Berhampur, Vigyanpuri, Ganjam, 760003, India}
\email{karabhijnan123@gmail.com}

\author{Pulkit Ojha}
\affiliation{Center for Theoretical Physics, Aleja Lotnikow 32/46, 02-668, Warszawa, Poland.}
\email{pojha@cft.edu.pl}

\author{Sudip Bhattacharyya}
\affiliation{Department of Astronomy and Astrophysics, Tata Institute of Fundamental Research, 1 Homi Bhabha Road, Colaba, Mumbai 400005, India}
\email{sudip@tifr.res.in}



\begin{abstract}
Sco X-1 is the brightest observed extra-solar X-ray source, which is a neutron star (NS) low-mass X-ray binary (LMXB), and is thought to have a strong potential for continuous gravitational waves (CW) detection due to its high accretion rate and relative proximity.
Here, we compute the long-term evolution of its parameters, particularly the NS spin frequency ($\nu$) and the surface magnetic field ($B$), to probe its nature and its potential for CW detection.
We find that Sco X-1 is an unusually young ($\sim 7\times10^6$~yr) LMXB and constrain the current NS mass to $\sim 1.4-1.6~{\rm M}_\odot$.
Our computations reveal a rapid $B$ decay, with the maximum current value of $\sim 1.8\times10^8$~G, which can be useful to constrain the decay models. 
Note that the maximum current $\nu$ value is $\sim 550$~Hz, implying that, unlike what is generally believed, a CW emission is not required to explain the current source properties.
However, $\nu$ will exceed an observed cut-off frequency of $\sim 730$~Hz, and perhaps even the NS break-up frequency, in the future, without a CW emission.
The minimum NS mass quadrupole moment ($Q$) to avoid this is $\sim (2-3)\times10^{37}$~g~cm$^2$, corresponding to a CW strain of $\sim 10^{-26}$. Our estimation of current $\nu$ values can improve the CW search sensitivity.

\end{abstract}



\section{Introduction} \label{sec:intro}
A neutron star (NS) low-mass X-ray binary (LMXB) is a binary stellar system, in which the NS accretes 
matter from a 
low-mass 
($\lesssim 1.5 M_\odot$) 
donor or donor star \citep{review_bhattacharya}. 
The accretion happens when the donor fills its Roche lobe.
Since this accreted matter has a large amount of specific angular momentum, it typically spins up the NS to a spin frequency ($\nu$) of a few hundred Hz. When the accretion finally stops, i.e., at the end of the LMXB phase, such rapidly spinning NSs can manifest as radio millisecond pulsars \citep[MSPs; ][]{Bhattacharyya_2021}.
Even in the LMXB phase, if the accretion disk is stopped by the NS magnetosphere, the disk matter may be channelled via accretion columns to the stellar magnetic poles. 
This can create azimuthally asymmetric emission from the spinning NS, and hence the observed X-ray emission can have periodic variation with the stellar spin frequency $\nu$. 
Such systems are called accretion-powered millisecond X-ray pulsars \citep[AMXPs; e.g.,][]{Patruno_amxp,Salvo2022}. 

If a spinning NS is asymmetric around its spin axis, implying a nonzero ellipticity ($\epsilon$) or mass quadrupole moment ($Q = I\epsilon$; $I$: NS moment of inertia), such an NS should emit gravitational waves continuously and spin down \citep{shapiro_book}. 
Note that other physical mechanisms, such as $r$ mode instability, $f$ mode instability, etc., can also cause continuous gravitational waves (CW) emission from NS \citep[for a comprehensive review, see][]{Glampedakis2018}. 
While such continuous waves (CW) have not been detected so far and only upper limits of $\epsilon$ have been estimated for a number of neutron stars \citep[e.g., ][]{Abbott_2017,Abbott_known1,Abbott_known2,Nieder_known1,Nieder_known2,sco_x1_gw2023}, such waves have been inferred from electromagnetic observations \citep[e.g.,][]{main_paper_sudip,Woan_2018,ellipcity_quadrupole,patruno_haskell_prl}.
Studies of CW from spinning NSs can be very useful to probe their physics, formation and evolution 
and hence the detection of CW is a holy grail of physics and astronomy.


What are the best sources to search for CW? 
Among NS LMXBs,
Scorpius X-1 (Sco X-1) is thought to be a very promising source for CW detection \citep[see][and references therein]{sco_x1_importance, Pagliaro_2023}
because of its relatively low distance \citep[$d = 2.8$ kpc; ][]{Bradshaw_1999} and a very high and persistent accretion rate $\dot{M}$ \citep[of the order of $10^{-8}~ \rm M_{\odot} \rm yr^{-1}$; ][]{sanchez_scox1,watts2008_scox1}.
Here is the justification for the latter argument.
Such a high $\dot{M}$ value could spin up the NS
beyond the break-up $\nu$ \citep[$\nu_{\rm break} \gtrsim 1200$~Hz; ][]{break_up}, which has not obviously happened.
Even if the current $\nu$ ($= \nu_{\rm curr}$) for Sco X-1 does not exceed $\nu_{\rm break}$, it could exceed an observationally indicated upper cut-off of $\nu$ \citep[$\sim 730$~Hz; ][]{deepto_nature,deepto_730, Patruno_2010NGW}.
These hint at a braking torque on the NS to keep its $\nu$ within a reasonable limit. 
CW can provide such a braking torque and cause the pulsar to spin down \citep[e.g., ][]{main_paper_sudip}.
However, in spite of this promise, the CW 
detection from Sco X-1 could be challenging and has not been detected yet in recent searches \citep{2021_gw_papa, sco_x1_gw2023, ScoX-1_2, Abbott_ScoX1}. This is because the source is not a pulsar and hence the $\nu$ value is not known \citep[See ][and references therein]{pulsation}.
Note that a known $\nu_{\rm curr}$ value reduces the volume of the parameter space and hence improves the sensitivity for CW searches.
However, in the absence of a measured $\nu_{\rm curr}$ value, one could estimate a range of $\nu_{\rm curr}$ values by the computation of long-term evolution of the LMXB and reproducing other known source parameters \citep[e.g., ][]{own_paper}.
Some such parameter values of Sco X-1 are orbital period $P_{\rm orb} = 0.787$~days \citep{Wang_2018}, donor-to-NS mass ratio $q = 0.52_{-0.15}^{+0.16}$ \citep{Wang_2018}, accretion rate $\dot M = 2.2\times10^{-8}~\rm M_{\odot}\rm yr^{-1}$ \citep{sanchez_scox1,watts2008_scox1} and donor star temperature $T_{\rm donor} < 4800$~K \citep{sanchez_scox1}.
Such a computation is also useful to probe the binary and stellar evolution and the NS physics and to estimate other unknown parameters, including $Q$ or $\epsilon$. 
However, while a few papers, such as \citet{chen_scox1, Van_2019}, computed the binary evolution for Sco X-1 to match some observationally known values of parameters (e.g., $P_{\rm orb}$, $q$, $\dot{M}$, $T_{\rm donor}$)
at the current time, they did not compute and constrain most NS parameters (e.g., $\nu$, $Q$).

\citet{own_paper} explored the NS LMXB evolution, including the evolution of NS and binary parameters, for ranges of initial and fixed parameter values.
Using the methods outlined in that paper,
here we compute the binary evolution for Sco X-1, 
and for the first time, calculate the $\nu$ evolution for this source considering different $Q$ values.
Note that, unlike some previous papers \citep[e.g., ][]{2021_gw_papa}, we do not consider the special condition of a balance between the accretion torque and the CW torque.
This condition, even if satisfied for certain periods, may not generally hold during the entire LMXB phase, and hence might not be valid at the current time.
This is because the mass transfer rate from the donor star evolves drastically throughout the LMXB phase \citep[e.g., ][]{own_paper}, and the CW may not entirely depend on accretion \citep[e.g., ][]{ellipcity_quadrupole}.
Therefore, we consider the general case of the mutually independent evolution of these torques depending on the evolution of various source parameter values.
Our computations constrain some NS parameters, e.g., mass ($M_{\rm NS}$), $\nu$, $\epsilon$, surface magnetic field ($B$), etc., which, among various aspects, could be useful for future CW detection from Sco X-1.

\section{Methods} \label{sec:methods}
We use MESA\footnote{\url{https://docs.mesastar.org/en/release-r23.05.1/}} (Modules for Experiments in Stellar Astrophysics), an open-source 1D stellar evolution code \citep{mesa2011,mesa2013,mesa2015,mesa2018,mesa2019}, to compute the binary evolution for Sco X-1. 
The binary parameter evolution and the Roche Lobe overflow (RLOF) primarily depend on the angular momentum ($J$) loss from the system, and the magnetic braking (MB) is the dominant mechanism for such a loss for the donor star of Sco X-1 \citep{Van_2019}.
However, a common prescription of MB \citep{rapapport_mb} cannot reproduce the observed high accretion rate of Sco X-1 and hence we use the ``Convection And
Rotation Boosted" (CARB) prescription with the loss of $J$ due MB given by \citep{Van_2019}:
\begin{equation}\label{carb mb}
\begin{split}
    \Dot{J}_{\rm MB}=&-\frac{2}{3}\dot{M}_{\rm W}^{-1/3} R^{14/3} \left( v_{\rm esc}^2 + 2 \Omega^2 R^2/K_2^2\right )^{-2/3}\\
    &\times\Omega_\odot\  B_{\odot}^{8/3}\ \left(\frac{\Omega }{\Omega_\odot}\right)^{11/3}\left(\frac{\tau_{\rm conv} }{\tau_{\odot, \rm conv}}\right)^{8/3}\ .
\end{split}
\end{equation}
Here, $\dot{M}_{\rm W}$ represents wind mass-loss rate from donor star, $R$ is the donor star radius, $v_{\rm esc}$ denotes surface escape velocity of the donor star, $\Omega$ is the rotation rate, $K_{2}$ represents limit where rotation rate has significant role and $\tau_{\rm conv}$ is convective turnover time.
The other components of $\dot J$ due to gravitational radiation and mass loss are considered according to their standard prescription mentioned in earlier papers \citep[e.g., ][]{main_jia&li,goodwin}.
However, although we primarily use the ``CARB" magnetic braking prescription, we also run MESA, following the general results of \citet{Yang_2024cboost}, for the ``Convection Boosted" (CBOOST) magnetic braking scheme \citep[Eq. 20][with $\xi = 2$]{cboost}. 
For baryonic to gravitational mass conversion for the NS, we use Eq. 19 of \citet{cipoletta} as outlined in \citet{own_paper}. We calculate NS radius as $R_{\rm NS} = AM_{\rm NS}^{1/3}$, where the value of the proportionality constant $A$ has been determined by taking an NS radius of 11.2 km for the NS mass of $1.4 ~\rm M_{\odot}$.

In order to model the NS spin evolution, we follow the torque prescription in \citet{main_paper_sudip} and references therein. The torques in the accretion and propeller phases ($N_{\rm acc}$ and $N_{\rm prop}$, respectively) are
\begin{equation}\label{Nacc}
N_{\rm acc} = \dot{M}\sqrt{GM_{\rm NS}r_{\rm m}} + \frac{\mu^2}{9r_{\rm m}^3}\Biggl[2\Bigl(\frac{r_{\rm m}}{r_{\rm co}}\Bigr)^3 - 6\Bigl(\frac{r_{\rm m}}{r_{\rm co}}\Bigr)^{\frac{3}{2}} + 3\Biggr],
\end{equation}
\begin{equation}\label{Nprop}
N_{\rm prop} = - \eta\dot{M}\sqrt{GM_{\rm NS}r_{\rm m}} - \frac{\mu^2}{9r_{\rm m}^3}\Biggl[3 - 2\Bigl(\frac{r_{\rm co}}{r_{\rm m}}\Bigr)^{3/2}\Biggr],
\end{equation}
where $r_{\rm m} (\propto B^{4/7}\dot{M}^{-2/7})$ and $r_{\rm co}(\propto \nu^{-2/3})$ are magnetospheric radius and corotation radius, respectively \citep[e.g., ][]{main_paper_sudip}, $\mu$ is the NS surface magnetic dipole moment and $\eta$ is a constant of the order of unity. 
In order to make our results realistic, we also consider that the NS surface magnetic field evolves as prescribed in \citet{shibazakiB}:
\begin{equation}\label{B eq}
    B(t) = \frac{B_{\rm i}}{1 + \Delta M_{\rm acc}(t)/m_{\rm B}},
\end{equation}
where $B_{\rm i}$ is the initial NS surface magnetic field, $B(t)$ is the NS surface magnetic field at time $t$ after the accretion of $\Delta M_{\rm acc}(t)$ mass, and $m_{\rm B}$ is a constant mass which determines the decay rate of the magnetic field.
The current value of the NS magnetic field is $B_{\rm curr}$.

Sco X-1 does not show pulsations (see section~\ref{sec:intro}) and it is clearly in the accretion phase \citep[i.e., $r_{\rm m} \leq r_{\rm co}$; ][]{main_paper_sudip} with a high accretion rate (see Figure~\ref{Sco x1 composite}).
Hence, at the current time, the NS magnetic field may be too weak to stop the accretion disk and cannot channel the accreted matter to the magnetic poles (see section~\ref{sec:intro}). 
Therefore, we consider that the accretion disk inner radius is currently either $R_{\rm NS}$ or $r_{\rm ISCO}$ (ISCO: innermost stable circular orbit), whichever is greater (i.e., max[$R_{\rm NS}$,$r_{\rm ISCO}$]; $r_{\rm ISCO} (= 6GM/c^2)$, $G$: gravitational constant, $c$: speed of light in vacuum). 
Thus, at the current time, the lack of pulsations in the accretion phase implies that $r_{\rm m} \ngtr {\rm max}[R_{\rm NS},r_{\rm ISCO}]$ and there is no significant disk-magnetosphere interaction.
Consequently, we consider 
$N_{\rm acc} = \dot{M}\sqrt{GM_{\rm NS}{\rm max}[R_{\rm NS},r_{\rm ISCO}]}$
(from Eq.~\ref{Nacc}) for the current time.

For the $Q > 0$ cases, we use the following CW torque 
\citep{Bildsten_1998}:
\begin{equation}\label{eq CW}
N_{\rm CW} = -\frac{32GQ^2}{5}\Bigl(\frac{2\pi\nu}{c}\Bigr)^5.
\end{equation}

In order to calculate the CW strain (which is crucial for CW detection), we use the equation \citep{gwapli_formula,Glampedakis2018}:
\begin{equation}\label{GW ampl}
h_0 = \frac{4\pi^2 G I}{c^4 d} f_{\rm CW}^2 \epsilon,
\end{equation}
where $f_{\rm CW}$ is the frequency of CW taken to be $2\nu$.

\section{Results} \label{sec:results}
\begin{figure*}
    \includegraphics[width=170mm, scale = 0.8]{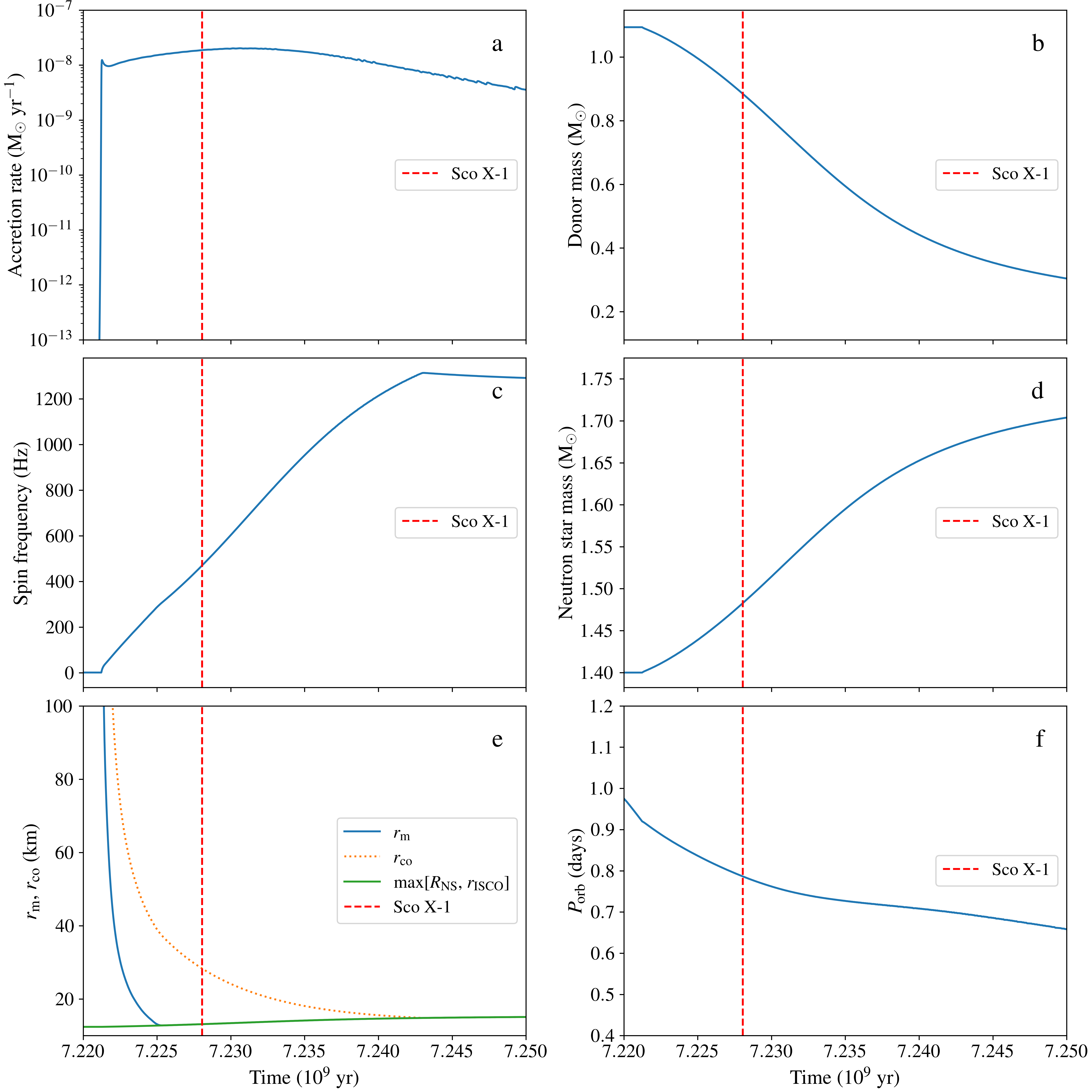}
\caption{Evolution of parameters (accretion rate [panel a], donor star mass [panel b], NS spin frequency [panel c], NS mass [panel d], characteristic radii [panel e], and binary orbital period [panel f]) for our best model of Sco X-1 evolution as computed using MESA (section \ref{sec:results}). 
The initial magnetic field ($B_{\rm i}$) is taken to be $6\times10^{11}$G with $m_{\rm B} = 1\times10^{-5}$.
The dashed vertical line shows the time at which all the observed parameters of Sco X-1, $P_{\rm orb}$, $\dot M$, $q$, $T_{\rm donor}$, match. 
The time mentioned on the x-axis is calculated from the start of the binary evolution computation.} 
\label{Sco x1 composite}
    
\end{figure*}

\begin{figure*}
\centering
\includegraphics[width=110mm, scale = 0.6]{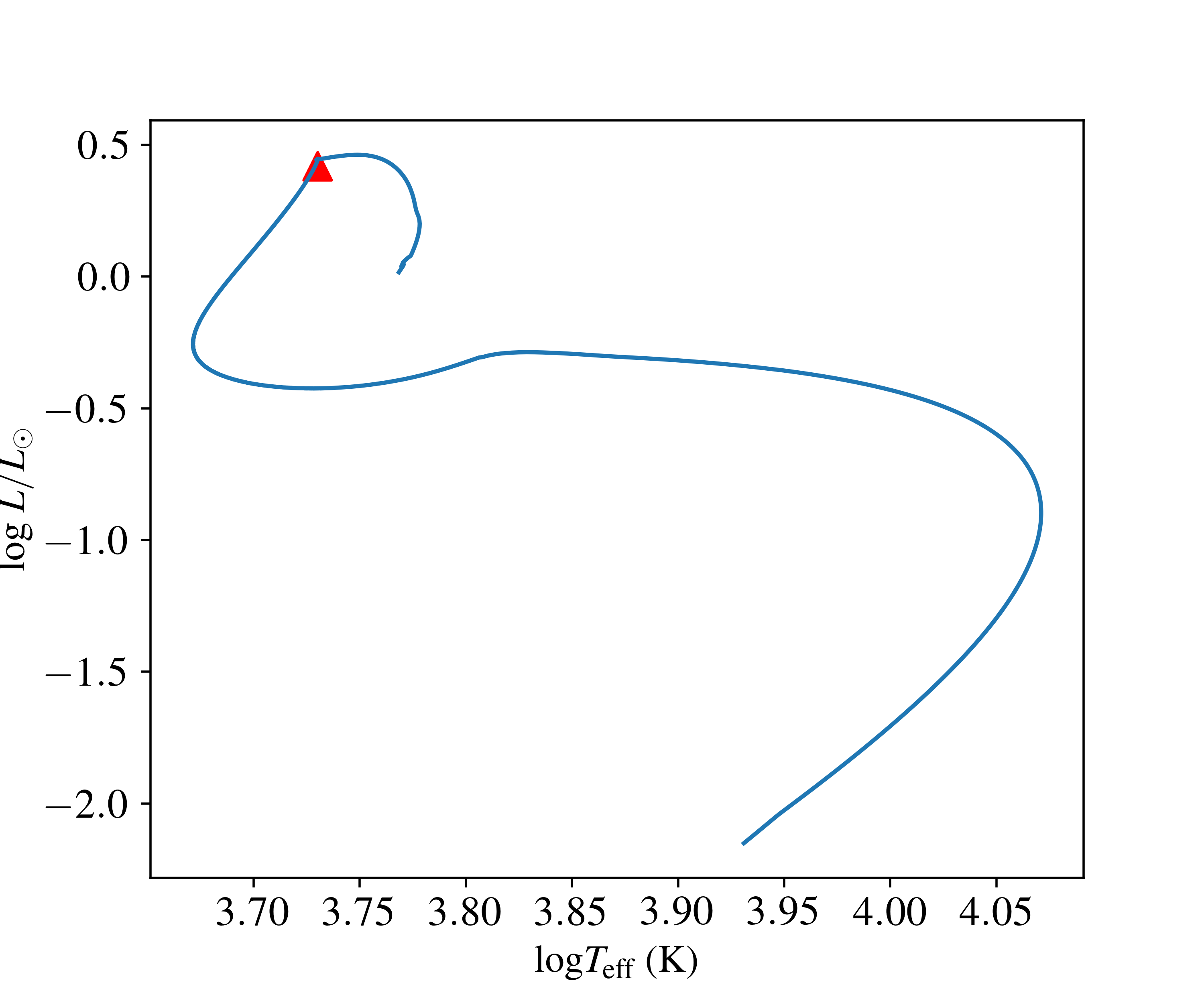}
\caption{The HR diagram for the donor star evolution (starting from the upper end) of Sco X-1, as computed with MESA. 
The initial parameter values are for the best model mentioned in section~\ref{sec:results}.
The current position of the source is depicted by a red triangle.}

\label{Sco_x1_hr}
\end{figure*}

\begin{figure*}
    \includegraphics[width= 180mm, scale = 2.2]{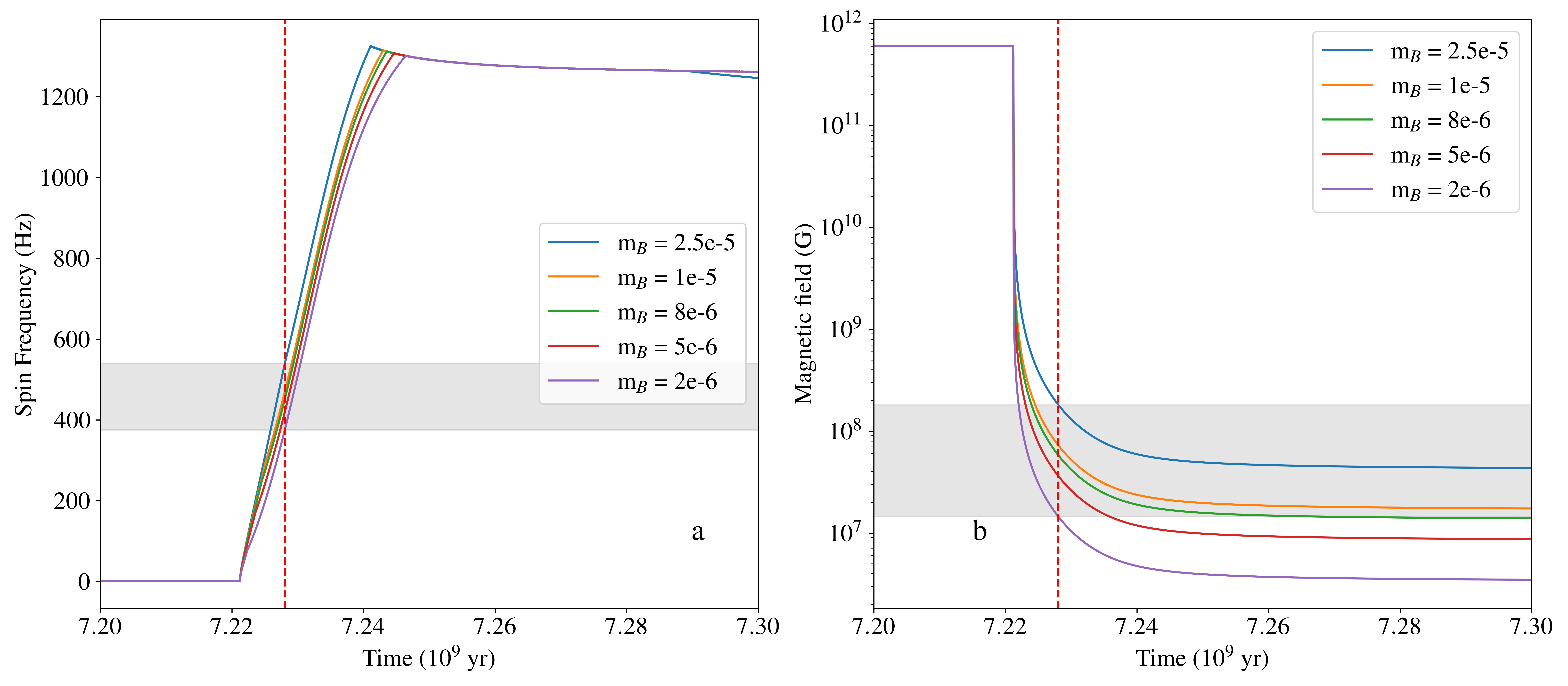}
    
\caption{The evolution of the spin frequency ($\nu$) [panel a] and the surface magnetic field ($B$) [panel b] of the NS of Sco X-1 for different values of $m_{\rm B}$ (Eq.~\ref{B eq}) considering the best model mentioned in section~\ref{sec:results}. 
The time mentioned on the x-axis is calculated from the start of the binary evolution computation. The accretion starts at $\sim 7.22\times10^{9}$ yr which is evident from the steep rise of the $\nu$-curve from its initial flat trajectory.
The initial $B$ is assumed to be $= 6\times 10^{11}$~G.
The dashed vertical line shows the time at which all the observed parameters of Sco X-1, $P_{\rm orb}$, $\dot M$, $q$, $T_{\rm donor}$, match.
The gray patches indicate the allowed ranges of the current $\nu~(= \nu_{\rm curr})$ and $B~(= B_{\rm curr})$ for the $m_{\rm B}$ values considered (see Section \ref{sec:results}).} 
\label{mb_spin}
    
\end{figure*}

We consider ranges of values for initial and fixed parameters \citep[e.g., ][]{own_paper}, such as donor star mass ($M_{\rm donor}$), $M_{\rm NS}$, $P_{\rm orb}$, fractional loss of the mass transferred from the donor 
star ($\beta$), irradiation efficiency ($\epsilon_{\rm irr}$), for our MESA  evolution models.
\citet{Van_2019} found that the reproduction of observed Sco X-1 parameter values requires a narrow initial parameter space.
Motivated by this, we consider initial ranges of $M_{\rm donor}$ ($1.0 - 1.2 ~ \rm M_{\odot}$), $P_{\rm orb}$ ($2.75 - 2.82$ days), $M_{\rm NS}$ ($1.35 - 1.6~ \rm M_{\odot}$), and also a range (0.3 - 0.7) of $\beta$ for our computation.
When we study the effects of change of one parameter value, we fix the other parameters at their canonical values (e.g., $M_{\rm NS}$ = 1.4 $\rm M_{\odot}$, $M_{\rm donor} = 1.1~\rm M_{\odot}$, $P_{\rm orb} = 2.80$ days, $\beta = 0.5$).
Note that the values of parameters, such as $\epsilon_{\rm irr}$, $\eta$, do not have visible effects for our computation to reproduce Sco X-1 parameter values.
By exploring the MESA evolution within the above-mentioned ranges, we find that the model with initial $P_{\rm orb} = 2.8$ days and initial $M_{\rm donor} = 1.1 ~\rm M_{\odot}$ reproduces the observed parameter values closely (Figure~\ref{Sco x1 composite}). 
We match the observed values of  $P_{\rm orb}$, $q$, $\dot M$ within $1\%$ and $T_{\rm donor}$ within $<~5 \%$, respectively. 
From this best model, we get $\dot M = 1.92\times 10^{-8}~ \rm M_{\odot} \rm yr^{-1}$ , $P_{\rm orb} = 0.786~\rm days$, $q = 0.596$, $T_{\rm donor}\sim 5000 ~\rm K$, all of which match most closely with the observed ranges of parameter values for Sco X-1 mentioned in Section \ref{sec:intro}. From the evolution, it can be seen that the accretion starts at $\sim 7.22\times10^9$ yr (Figure \ref{Sco x1 composite}, \ref{mb_spin}).
Moreover, note that the best model implies a current value of $M_{\rm NS}$ of $\sim1.48~\rm M_{\odot}$, and considering all reasonably matching evolution models, the current $M_{\rm NS}$ range is  $\sim1.42 - 1.6~\rm M_{\odot}$.
Figure~\ref{Sco_x1_hr} shows the Hertzsprung–Russell (HR) diagram of the evolution of donor star of Sco X-1 and its current position.

With our best model, we calculate the spin evolution and estimate the current $\nu$ value ($\nu_{\rm curr}$) of Sco X-1 \citep[see ][and section~\ref{sec:methods} for details]{own_paper}. 
For the default model, we keep $Q = 0$, i.e., we take a case without CW. 
However, with the usually considered value of $m_{\rm B}$ \citep[$\sim 10^{-4} ~\rm M_{\odot}$; ][]{shibazakiB} in Eq.~\ref{B eq}, we cannot satisfy the condition of no pulsation in the accretion phase, viz., $r_{\rm m} \ngtr {\rm max}[R_{\rm NS},r_{\rm ISCO}]$ (see section~\ref{sec:methods}).
Hence, we explore a range of lower $m_{\rm B}$ values which allows a faster decay of NS magnetic field (Figure~\ref{mb_spin}), and, in turn, provides lower $B_{\rm curr}$ values resulting in lower $r_{\rm m}$ values \citep[e.g., ][]{ main_paper_sudip}.
We assume a reasonable lower limit of $B_{\rm curr}$ \citep[$\sim 10^{7}$ G; see ][]{amxpmagDB}.

The conditions $r_{\rm m} \ngtr {\rm max}[R_{\rm NS},r_{\rm ISCO}]$ and $B_{\rm curr} \gtrsim 10^{7}$~G provide a range of $m_{\rm B}$ for each $B_{\rm i}$.
We consider a reasonable $B_{\rm i}$ range of $6\times10^{11}-2\times10^{12}$~G \citep[e.g., ][]{review_bhattacharya}.
This provides an overall range of $m_{\rm B}$ ($\sim 6\times10^{-7}-2.5\times10^{-5}$; for our best model; Table~\ref{table1}). 
Note that, due to the above two conditions, $B_{\rm curr}$ has the similar range for all $B_{\rm i}$ values (upper value of $B_{\rm curr} \sim 1.8\times10^8$~G; see Table~\ref{table1}, for example).
Moreover, as indicated above, a lower $m_{\rm B}$ value implies a lower $B_{\rm curr}$ value (Table~\ref{table1} and Figure~\ref{mb_spin}).
Besides, as $B$ decays faster for a lower $m_{\rm B}$ value, the spin-up torque also decreases faster (see section~\ref{sec:methods}; Eq.~\ref{Nacc}), and hence the spin-up rate is slower.
Therefore, at a given time (e.g., the current time), $\nu$ is lower for lower $m_{\rm B}$ and $B$ values (see Table~\ref{table1} and Figure~\ref{mb_spin}). 
However, as the range of $B_{\rm curr}$ does not depend on $B_{\rm i}$, the range of $\nu_{\rm curr}$ ($\sim 370-540$~Hz for the best model) also does not depend on $B_{\rm i}$ (Table~\ref{table1} and Figure~\ref{mb_spin}).
Furthermore, we consider other evolution models, which reasonably match the Sco X-1 observed parameter values, and get similar ranges of $m_{\rm B}$, $B_{\rm curr}$ and $\nu_{\rm curr}$, as mentioned above. From all of these models combined, $\nu_{\rm curr}$ comes out to be $\sim 350-550$~Hz.

Let us now consider the effects of CW ($Q > 0$) on the NS spin evolution for Sco X-1.
For the LMXB phase and the radio MSP phase (section~\ref{sec:intro}), $\epsilon \sim 10^{-9}$ or $Q \sim 10^{36}$~g~cm$^2$ for NSs has been inferred using electromagnetic observations
\citep[e.g., ][]{Woan_2018,ellipcity_quadrupole}.
Nevertheless, we explore a $Q$ range of $\sim 10^{36-38}$~g~cm$^2$ ($\epsilon \sim 10^{-9}-10^{-7}$), and even higher.

The effects of $Q$ on the NS spin evolution for Sco X-1 are shown in Figure~\ref{Q}. Here, we consider a $Q$ range of $\sim 10^{36-38}$~g~cm$^2$,
where the lower $Q$ values are motivated by \citet{ellipcity_quadrupole}.
Note that $Q \ge 2.3\times10^{37}$~g~cm$^2$ for our best model (section~\ref{sec:results}), $B_{\rm i} = 6\times10^{11}$~G and $m_{\rm B} = 2.5\times10^{-5}$
imply that $\nu$ did not and will not exceed the  observationally indicated upper cut-off of $\nu$ ($\sim 730$~Hz; section~\ref{sec:intro}) throughout the LMXB phase of Sco X-1.
This minimum $Q$ is called $Q_{\rm min(730)}$ for the above-mentioned model. If we consider a very high value of $Q$ (e.g., $10^{40}$~g~cm$^2$) as the $Q_{\rm max}$, then the allowed $\nu_{\rm curr}$ range for $Q_{\rm min(730)}-Q_{\rm max}$ for the above-mentioned model is $\sim 500-70$~Hz (Figure~\ref{Q}).
For all the matching evolution models and the allowed $m_{\rm B}$ range for Sco X-1, the allowed $\nu_{\rm curr}$ range for $Q_{\rm min(730)}-Q_{\rm max}$ is $\sim510-60$~Hz.
However, for a $Q$ range of $0-Q_{\rm max}$, the range of $\nu_{\rm curr}$ for all matching models is $\sim 550-60$~Hz.

We calculate the current CW strain, $h_{\rm 0,curr}$, as a function of $Q$ for Sco X-1 using Eq.~\ref{GW ampl}.
In order to constrain $h_{\rm 0,curr}$, we plot $h_{\rm 0,curr}$ versus $Q$ in Figure~\ref{Q}b.
In the same panel, we also plot $\nu_{\rm curr}$ versus $Q$.
Let us understand these two curves.
As one gradually considers higher $Q$ values starting from a low value, initially $\nu_{\rm curr}$ does not decrease much when the $Q$ values are still small (as $N_{\rm CW} \propto Q^2$; Eq.~\ref{eq CW}). 
Consequently, $h_{\rm 0,curr}$ increases rapidly with $Q$ (roughly proportionally; Eq.~\ref{GW ampl}; see Figure~\ref{Q}b).
But, for larger $Q$ values ($\gtrsim 10^{38}$~g~cm$^2$), $\nu_{\rm curr}$ decreases rapidly (as $N_{\rm CW} \propto Q^2$; Eq.~\ref{eq CW}). 
As a result, $h_{\rm 0,curr}$ increases slowly with $Q$, because $h_{\rm 0,curr} \propto \nu_{\rm curr}^2Q$ (Eq.~\ref{GW ampl}).
This means, even if the $Q$ value of the NS in Sco X-1 is much larger than $\sim 10^{38}$~g~cm$^2$, the $h_{\rm 0,curr}$ value is not significantly greater than the value for $Q \sim 10^{38}$~g~cm$^2$.
In fact, we do not expect an $h_{\rm 0,curr}$ value greater than a few times $10^{-26}$ (see Figure~\ref{Q}b).
This points to the requirement of a minimum instrument capability for the CW detection from Sco X-1.

For the models with the magnetic braking prescription ``CBOOST" (see section~\ref{sec:methods}), we use the similar initial parameter values around the mentioned ranges for the ``CARB" model. 
With this prescription too, the observed parameter values for Sco X-1 can be closely matched, except for $P_{\rm orb}$ which could be matched within $\sim 20~\%$ at the best.
The NS parameter (such as $\nu$, $B_{\rm curr}$) values calculated for ``CBOOST" are similar to those obtained for the ``CARB" models. 
For example, $\nu_{\rm curr}$ values are within the range of $\sim 300 - 600$ Hz, which overlap with the range obtained for``CARB".

\begin{table*}[htbp]
\begin{center}
\caption{Several example initial magnetic field ($B_{\rm i}$) values and their corresponding $m_{\rm B}$, current magnetic field ($B_{\rm curr}$) and current spin frequency ($\nu_{\rm curr}$) ranges for Sco X-1 using evolution computations with MESA (for no CW; sections~\ref{sec:methods} and \ref{sec:results}).}

\begin{tabular}{|l|c|c|c|l|} 
    \hline
    $B_{\rm i}$(G) & $m_{\rm B}$ & $B_{\rm curr}$(G) & $\nu_{\rm curr}$ (Hz) \\
    \hline
    \multirow{3}{*}{$6\times 10^{11}$}  
        & $2.5\times10^{-5}$ & $1.81\times 10^8$  & 539.8 \\
        & $8\times10^{-6}$ & $5.8\times10^7$ & 451.8  \\
        & $5\times10^{-6}$ & $3.62\times10^7$ & 419.8  \\
        &$2\times10^{-6}$& $1.45\times10^7$& 375.7\\
    \hline
    \multirow{3}{*}{$8\times10^{11}$}  
        & $1.8\times10^{-5}$  & $1.74\times10^8$ & 537.7  \\
        & $8\times10^{-6}$  & $7.74\times10^7$ & 474.6 \\
        & $4\times10^{-6}$  & $3.87\times10^7$ & 423.6 \\
        & $1.5\times10^{-6}$& $1.45\times10^7$& 375.7\\
    \hline
    \multirow{2}{*}{$1\times10^{12}$}  
        & $1.5\times10^{-5}$  & $1.81\times10^8$ & 539.8 \\
        & $8\times10^{-6}$  & $9.67\times10^7$ & 492.2\\
        &$4\times10^{-6}$ & $4.83\times10^7$& 438.6\\
        & $1.2\times10^{-6}$& $1.45 \times10^7$ & 375.7\\
    \hline
    \multirow{2}{*}{$2\times10^{12}$}  
        & $7.5\times10^{-6}$  & $1.81\times10^8$ & 539.7\\
        & $4\times10^{-6}$  & $9.67\times10^7$ & 492.3\\
        &$1\times10^{-6}$ & $2.41\times10^7$& 397.2\\
        & $6\times10^{-7}$& $1.45 \times10^7$ & 375.6\\
    \hline
    
\end{tabular}
\label{table1}
\end{center}
\end{table*}


\section{Discussion and conclusion} \label{sec:discussion}
Sco X-1 is the first discovered and the 
brightest observed extra-solar X-ray source.
This persistent source is thought to be among the best potential sources to detect CW (section~\ref{sec:intro}).
Here, we compute the long-term evolution of this source to understand its nature and its potential for CW detection.
Our computation of evolution of Sco X-1 using the MESA code shows that it is an unusually young NS LMXB (age $\sim 7\times10^6$~yr) with a predicted total LMXB lifetime of $\sim 5\times10^8$~yr, which is also surprisingly short 
(see appendix).
The LMXB phase ends when the donor star does not fill its Roche lobe anymore.
Moreover, the current accretion rate of the NS of Sco X-1 is rather high, which cannot be explained with a usual magnetic braking prescription and requires a boosted braking prescription, such as ``CARB" or ``CBOOST".
While our evolution computations reproduce the known parameter values of Sco X-1, we could also constrain the current NS mass to the range $\sim1.42 - 1.6~\rm M_{\odot}$
(section~\ref{sec:results}). 
Note that mass measurement of an NS is important to probe the nature of the degenerate NS core matter at  supra-nuclear densities \citep[][]{sudip_adspr}.

\begin{figure*}[h]
\centering
\includegraphics[width=180mm,scale=2.2]{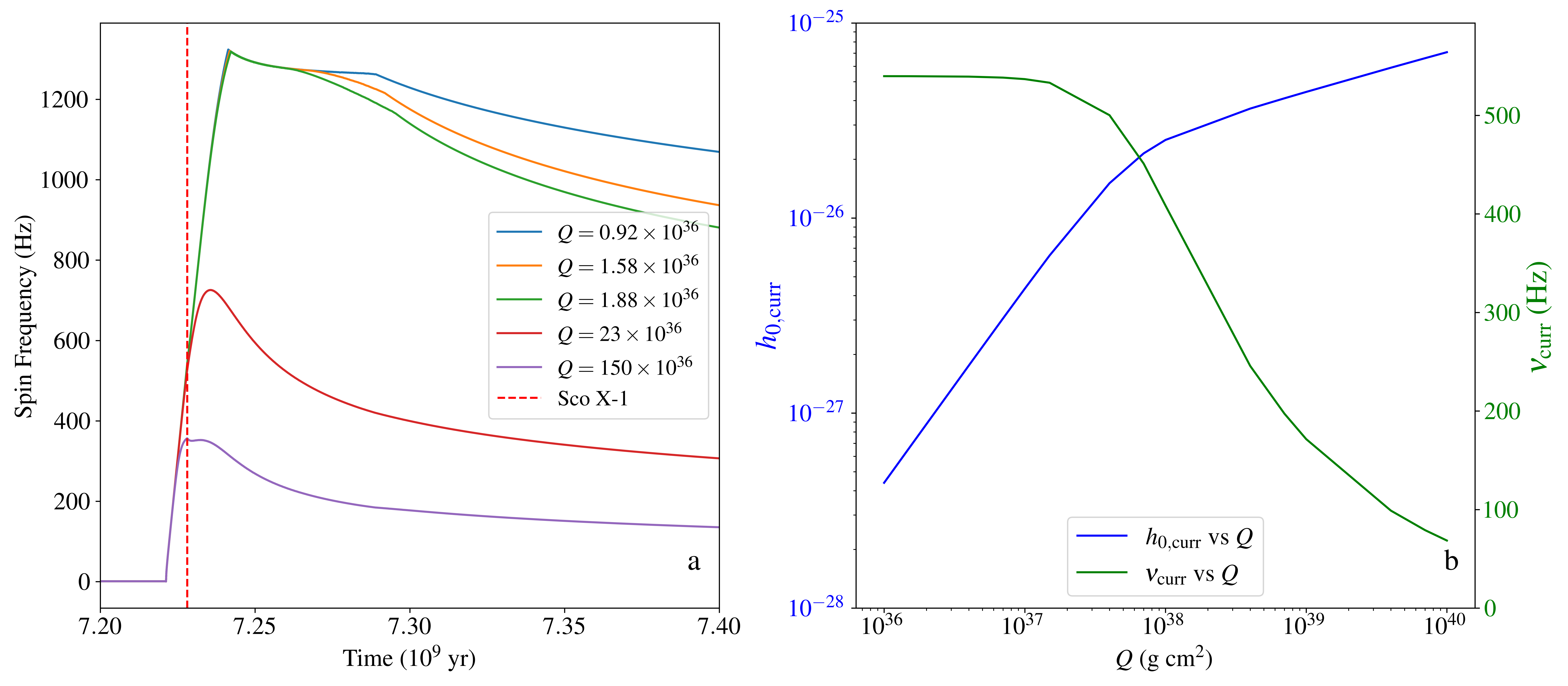}
\caption{{\it Panel a:} The evolution of the NS spin frequency ($\nu$) for Sco X-1 for different values of $Q$ (Eq.~\ref{eq CW}) considering the best model mentioned in section~\ref{sec:results}. 
The time mentioned on the x-axis is calculated from the start of the binary evolution computation. 
The $B_{\rm i}$ and $m_{\rm B}$ values are $6\times10^{11}$ G and $2.5\times10^{-5}$, respectively.
The dashed vertical line shows the time at which all the observed parameters of Sco X-1, $P_{\rm orb}$, $\dot M$, $q$, $T_{\rm donor}$, match.
{\it Panel b:}
The CW strain amplitude $h_{\rm 0, curr}$ and the NS spin frequency $\nu_{\rm curr}$ for Sco X-1 at the current time for a range of $Q$ values (for the above-mentioned best model, $B_{\rm i}$ and $m_{\rm B}$ values; section~\ref{sec:results}).
Note that $h_{\rm 0, curr}$ increases slowly beyond $Q \sim 10^{38} \rm g~ \rm cm^{2}$ ($\epsilon \sim 10^{-7}$).}
\label{Q}
\end{figure*}

We compute the NS spin frequency $\nu$ and surface magnetic field $B$ evolution for Sco X-1 for the first time to the best of our knowledge. 
Here, we utilize the lack of observed pulsations from this source (see sections~\ref{sec:intro} and \ref{sec:methods}) for the first time.
We find that the maximum value of  the current $B$, i.e., of $B_{\rm curr}$ is $\sim 1.8\times10^8$~G.
Thus, it is remarkable that the $B$ value of the NS decays by at least $\sim 4$ orders of magnitude in just $\sim 7$ million years, and this fast $B$ decay is also supported by our preferred low $m_{\rm B}$ values (section~\ref{sec:results}).
Note that the physics of the $B$ decay of NSs is a topical problem of astrophysics \citep{Mag_evolution_universe}, and our finding should be important to constrain models.
For example, the high $\dot{M}$ could screen the magnetic field lines so well that they could not re-emerge easily. 
Alternatively, the Ohmic decay could be more efficient due the high temperature because of the high $\dot{M}$.
It is important to note that, even without CW, the current $\nu$, i.e., $\nu_{\rm curr}$, value is not more than $\sim 550$~Hz (section~\ref{sec:results}). 
This value is well below both the break-up $\nu$ and the observed cut-off $\nu$ (see section~\ref{sec:intro}).
This means that CW from Sco X-1 is not required considering its current properties.
Hence, unlike what is usually believed (see section~\ref{sec:intro}), the current high $\dot{M}$ of this source does not need CW, and hence Sco X-1 might not be one of the best sources for CW detection.
Nevertheless, the source could still emit CW, and considering a $Q$ value up to $\sim 10^{40}$~g~cm$^2$, the lower limit of $\nu_{\rm curr}$ could be $\sim 60$~Hz. Note that our estimation of $\nu_{\rm curr}$ values as a function of $Q$ can be very useful to improve the sensitivity for CW searches (see section~\ref{sec:intro}).
However, our evolution computations show that the $\nu$ value will exceed $\sim 1200$~Hz in the future, if no CW is considered (see appendix).
Thus, if we assume that $\nu$ will not exceed the observed cut-off $\nu$ (see section~\ref{sec:intro}) even in the future, then we need a $Q$ of at least $\sim (2-3)\times10^{37}$~g~cm$^2$, implying a current 
CW strain $h_0$, i.e., $h_{\rm 0,curr}$ of $\sim 10^{-26}$ (see section~\ref{sec:results}, Figure~\ref{Q}b).
However, even for a very high $Q \sim 10^{40}$~g~cm$^2$ (i.e., $\epsilon \sim 10^{-5}$), $h_{\rm 0,curr}$ would not be more than  $\sim 7\times10^{-26}$ (see section~\ref{sec:results}, Figure~\ref{Q}b).
These upper limits on $h_{0,\rm curr}$ are somewhat comparable to those obtained from the  Advanced LIGO data \citep[e.g,][]{Abbott_2017,sco_x1_gw2023}.



%



\clearpage

\bibliography{sample631}{}
\bibliographystyle{aasjournal}

\appendix
Here, we show, in Figure~\ref{Sco x1 full}, the evolution of some  parameters, performed by the MESA code (see section~\ref{sec:methods}), for the entire LMXB phase of Sco X-1. 
This is useful to understand the nature of this source, and thus, to provide a broad and general perspective of our results and conclusions.

\renewcommand{\thefigure}{A\arabic{figure}}
\setcounter{figure}{0}
\begin{figure*}
    \includegraphics[width=170mm, scale = 0.8]{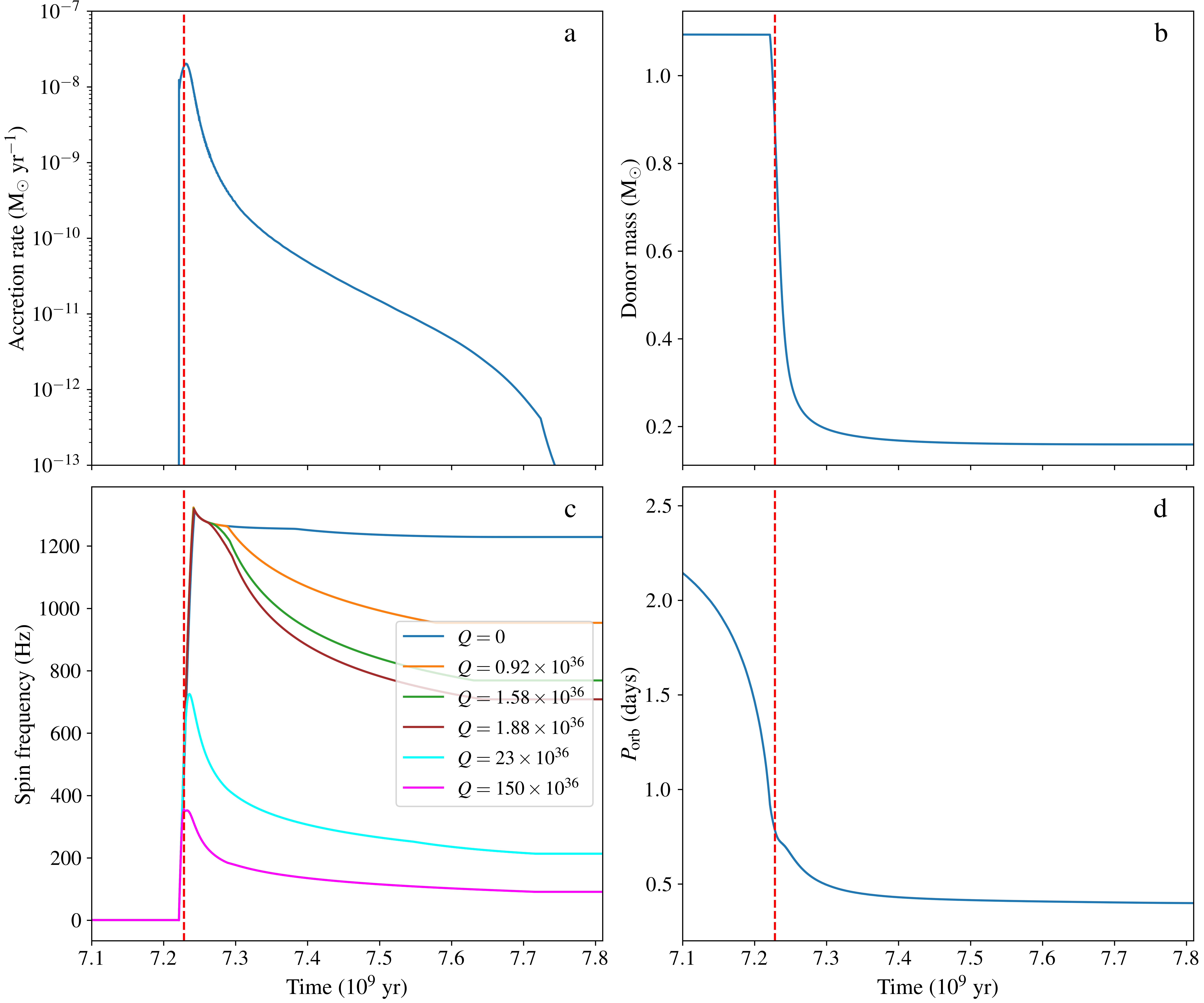}
\caption{Evolution of parameter values, accretion rate (panel a), donor star mass (panel b), NS spin frequencies ($\nu$ for various $Q$ values; panel c), and binary orbital period ($P_{\rm orb}$; panel d) during the entire LMXB phase for the best model of Sco X-1 (see section~\ref{sec:results}). 
$B_{\rm i}$ and $m_{\rm B}$ values are $6\times10^{11}$~G and $2.5\times10^{-5}$, respectively.
The time mentioned on the x-axis is
calculated from the start of the binary evolution computation, and the dashed vertical line shows the time at which all the observed parameters of Sco X-1, $P_{\rm orb}$, $\dot M$, $q$, $T_{\rm donor}$, match.
This figure shows that Sco X-1 is a relatively young NS LMXB and its LMXB phase will be relatively short-lived.}
\label{Sco x1 full}

\end{figure*}




\end{document}